\documentclass{llncs}

\newif\ifdraft
\draftfalse

\usepackage[T1]{fontenc}
\usepackage[utf8]{inputenc}
\usepackage{textcomp}

\ifdraft
\usepackage[show]{ed}
\usepackage{pdfsync}
\else
\usepackage[hide]{ed}
\usepackage{microtype}
\fi

\usepackage{paralist}
\usepackage{courier}
\usepackage{amsmath}
\usepackage{amstext}
\usepackage{graphicx}
\usepackage{fancyvrb}
\usepackage{relsize}

\usepackage{wasysym}

\def\thetitle{wiki.openmath.org -- how it works, how you can participate}

\usepackage[pdftex,pdfstartview=FitV,plainpages=false,pdfpagelabels,colorlinks=true,hypertexnames=true]{hyperref}
\hypersetup{%
  pdfauthor = {Christoph Lange},%
  pdftitle = {wiki.openmath.org – how it works, how you can participate},%
  pdfkeywords = {OpenMath wiki SWiM collaboration MKM content dictionaries semantic
    markup}%
}
\usepackage{hyperref}

\title{\thetitle} \author{Christoph Lange} \institute{Computer Science, Jacobs University
  Bremen, \email{ch.lange@jacobs-university.de}}

\begin{document}

\maketitle

\begin{abstract}
  At \url{http://wiki.openmath.org}, the OpenMath 2 and 3 Content Dictionaries are
  accessible via a semantic wiki interface, powered by the SWiM system.  We shortly
  introduce the inner workings of the system, then describe how to use it, and conclude
  with first experiences gained from OpenMath society members working with the system and
  an outlook to further development plans.
\end{abstract}

\section{Introduction: The OpenMath Content Dictionaries}
\label{sec:om3}

OpenMath~\cite{Openmath:web} is a semantic markup language (``content markup language'')
for mathematical formulæ that originated as a shared knowledge representation for
applications in computer algebra and automated theorem proving in the mid-1990s and got
further applied in areas as diverse as e-learning, scientific publishing, and interactive
geometry.  OpenMath defines an abstract data model for representing mathematical objects
and two concrete syntaxes for it, a binary and a more common XML one.  Important building
blocks of mathematical objects are numbers, variables, symbols, and applications of
mathematical objects to other mathematical objects.  Any concrete operator, constant, set,
or function can be a symbol.  In contrast to, e.\,g., earlier versions of MathML, the
symbol supply of OpenMath is constantly increasing due to its extensibility by so-called
\emph{content dictionaries} (CDs).

\begin{figure}
\RecustomVerbatimEnvironment{Verbatim}{Verbatim}{fontsize=\footnotesize,commandchars=\\\{\},codes={\catcode`$=3\catcode`$=3}}
\begin{Verbatim}
<CDDefinition>
<Name>plus</Name>
<Role>application</Role>
<Description>The symbol representing an n-ary commutative 
 function plus.</Description>
<CMP>for all a,b | a + b = b + a </CMP>
<FMP>$\beta(\operatorname{quant1\#forall},a,b,$
         $@(\operatorname{relation1\#eq},@(\operatorname{arith1\#plus},a,b),@(\operatorname{arith1\#plus},b,a)))$</FMP>
</CDDefinition>
\end{Verbatim}
\caption{Definition of the \textit{arith1\#plus} symbol}
\label{fig:cddef}
\end{figure}

A CD is a collection of (usually closely related) mathematical symbols, each with a
\emph{name} and a mandatory informal \emph{description} (cf.\ fig.~\ref{fig:cddef}).
Further information about symbols is optional but recommended to have: mathematical
properties of the symbol, both in a formal (FMP) and an informal (``commented'', CMP)
flavour, and examples of applying the symbol.  The language for expressing this
information is part of the OpenMath standard.  Besides the proper CD file (named e.\,g.\
\texttt{number-theory.ocd}), there can be additional files: OpenMath does not commit to a
particular \emph{type system}, so it allows for types of symbols to be specified in
separate files parallel to the CD, one per type system.  The most common type system in
the OpenMath community is, however, Davenport's Small Type System
(STS~\cite{Davenport:stso99}); types in that system would be given in a file named
\texttt{number-theory.sts}.

Furthermore, there is no doubt that \emph{notations} must be specified for symbols in some
way, if OpenMath objects should ever be presented to a human reader, but opinions diverge
on whether this should be done in CD-like files or not.  David Carlisle and others believe
that directly writing XSLT (one file per CD, one template per symbol) does a good job in
transforming OpenMath to Presentation MathML.  The advantage of XSLT is its expressive
power (it's Turing-complete!), which comes at the expense of human comprehensibility,
though.  Paul Libbrecht and Michael Kohlhase (of whose ``camp'' the author is a member)
thus prefer CD-like dictionaries of XML-based notation definitions in a more compact
syntax.  They believe that, given a sufficient support for pattern matching or declarative
symbol$\mapsto$notation mappings, most, if not all aspects of mathematical notation can be
handled, and authored much more intuitively.  Libbrecht et al.\ \emph{generate} XSLTs from
notation definitions that use pattern matching, whereas Kohlhase et al.\ have implemented
a dedicated renderer (actually two ones, which are being merged) that directly renders
mathematical objects using either declarative or pattern-based notation
definitions~\cite{KohLanRab:pmcfe07,TR:KLMMR:NfAD,KMR:NoLMD08}.

\section{Authoring and Reviewing OpenMath CDs}
\label{sec:author-review}

While everybody is free do define his own CDs for his purposes, the OpenMath Society
maintain a collection of \emph{official} CDs~\cite{JDPL:TFtEOaiU} that have undergone a
review process.  Still, the content of an official CD is not fixed: It might still contain
mistakes that have slipped through the review, or there might be ways to improve the
informal descriptions of symbols, or relevant mathematical properties and examples to
add.

As said in the introduction, one CD is essentially a file -- containing several metadata
fields on top, and then one \textit{CDDefinition} block per symbol.  The official CDs are
maintained in a Subversion repository at \url{https://svn.openmath.org}.  Developers
participating in their maintenance check out a working copy of that repository, edit the
CD files locally with a text or XML editor, and then commit their changes.  RIACA have
developed a Java-based CD editor~\cite{URL:riaca-openmath}, the only CD editor besides
ours that we are aware of.  The RIACA CD editor, however, rather focuses on generating
Java code for programs dealing with OpenMath objects from CDs than on CD maintenance, and
its development  seems to have been discontinued for at least three years.

Issues with the CDs are usually being discussed on the OpenMath mailing list
(\url{om@openmath.org}) in case of fixing bugs in existing CDs, or on the OpenMath 3
mailing list (\url{om3@openmath.org}) in case of the overhaul of the CDs and alignment
with the Content MathML specification for the upcoming OpenMath
3~\cite{DavKoh:umoattsea09}.  As an alternative for OpenMath 3, there is an installation
of the Trac issue tracking system (cf.~\cite{trac:web}) at
\url{https://trac.mathweb.org/OM3}.

For presenting a CD to human readers, the elements of the OpenMath CD language are usually
transformed to the desired output format (most commonly XHTML) using XSLT, and the
OpenMath objects occurring inside the FMPs and examples are rendered as described in
section~\ref{sec:om3}.  This presentation process is usually controlled by makefiles.

\subsection{Three CD Editing Use Cases}
\label{sec:use-cases}

In the remainder of this paper, I will focus on supporting three common use cases.  First,
the traditional way of handling these cases will be presented, to pave the way for showing
how they are handled in the OpenMath wiki.

\subsubsection{Minor Edits:}
\label{sec:uc-minoredit}

Fixing minor mistakes does not change the semantics of a symbol.  Consider correcting a
spelling mistake in a description, or renaming a bound variable in a mathematical object
that does not occur as a free variable in a subexpression.  Supported by a text or XML
editor only, which is not aware of the particular features of OpenMath CDs, such a fix
would be done as follows (assuming that the mistake is in a CD from \url{openmath.org}):
\begin{enumerate}
\item Update the working copy of the OpenMath CDs
\item Open the CD file in question
\item Navigate to the \textit{Description} child of the symbol in question
\item Fix the mistake
\item Commit the file (and, ideally: commit that file only, and give a meaningful log
  message that exactly refers to the symbol where the mistake was fixed)
\end{enumerate}

\subsubsection{Discussing and Implementing Revisions:}
\label{sec:uc-argu}

Major revisions that change the semantics of a symbol have to be discussed among the
developers before implementing them.  Usually, the discussion starts with pointing out a
problem (e.\,g.\ an FMP for a concrete symbol is wrong or misleading).  Let us assume that
the developer who identified the problem does not know how to solve it.  Then, he would
have to make others aware of the problem, e.\,g.\ by an e-mail to the OpenMath mailing
list.  Pasting a link to the Subversion URL of the CD in question into that e-mail helps
others to inspect the problematic part\footnote{Trac features a more immediate and
  comprehensive integration of a trouble ticket system with a Subversion repository, but
  that is not currently possible for OpenMath, as the Trac and the Subversion repository
  are running on different servers.}.  Other developers would then reply to this e-mail
and propose solutions, and again by replying to their mails, the solutions would be
discussed, until the community agrees on one to be implemented.

\subsubsection{Editing and Verifying Notations:}
\label{sec:uc-notation}

Suppose that an example or FMP for a symbol $\sigma$ in one CD uses a symbol $\tau$ from
another CD and that the notation defined for $\tau$ is wrong.  Concretly, imagine $\sigma$
being the cumulative distribution function of the normal distribution, $\tau$ the integral
symbol occurring in the definition of $\sigma$, and then imagine that the formatting of
its lower and upper bounds is wrong.  Here is how an author would fix this:
\begin{enumerate}
\item Identify the formal symbol name and CD of $\tau$
\item\label{it:find-notdef} Navigate to the file where the notation of $\tau$ is defined
\item Try to fix the notation definition
\item Regenerate the human-readable presentation of the CD defining $\tau$ (and,
  ideally: regenerate \emph{all} CD presentations where $\tau$ occurs)
\item Open the regenerated presentation and check if it is correct (if not, back
  to~\ref{it:find-notdef})
\item Commit the file containing the notation definition, giving a meaningful log message
\end{enumerate}

\section{The OpenMath Wiki}
\label{sec:wiki}

\begin{figure}
\includegraphics[width=\textwidth,height=9.1cm]{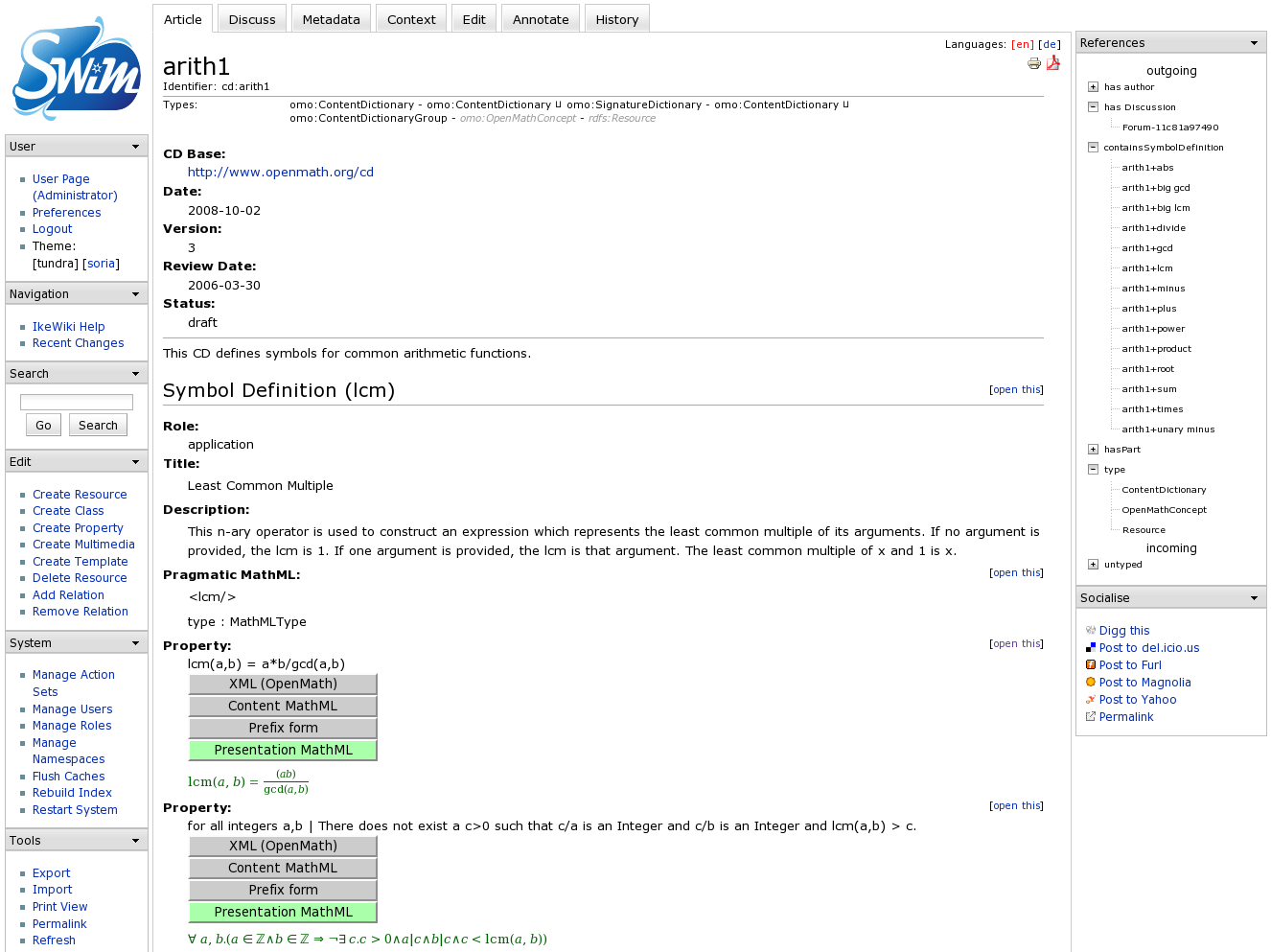}
\caption{An OpenMath CD in SWiM.  Notice the navigation links on the right side.}
\label{fig:swim}
\end{figure}

From the previous use case descriptions it evident that a better tool support is needed to
aid maintenance of the OpenMath CDs.  SWiM is a wiki -- a system for collaboration on
knowledge collections on the web --, a \emph{semantic wiki for mathematics} in
particular~\cite{lange:swim-demo08}.  It aims at offering intelligent collaboration
services to authors of mathematical documents in semantic markup languages -- such as
OpenMath CDs.  SWiM's notion of ``semantics'' is restricted to decidable structural
aspects of documents and CDs; it does not capture the full semantics of OpenMath objects.
Having presented first ideas at the OpenMath workshop in January
2008~\cite{Lange:SWiM-OpenMath08}, the author decided to further pursue supporting the
OpenMath CD review as a case study for SWiM and set up an instance of the system at
\url{http://wiki.openmath.org} in September 2008.  Figure~\ref{fig:swim} shows a CD in the
browsing view of SWiM.  In the remainder of this section, it will be discussed how SWiM
supports the use cases introduced in section~\ref{sec:use-cases}.

\subsection{Minor Edits}
\label{sec:swim-minoredit}

We have identified three different types of knowledge in OpenMath CDs: the structural
outline of a CD (e.\,g.\ defining what symbols a CD defines), metadata (of such structural
units, e.\,g.\ their informal descriptions or the date of revision), and OpenMath objects
(inside FMPs and examples).  For each of them, SWiM offers a dedicated editor
(see~\cite{LangeGonzalez:SWiM-Sentido08}) for details.

It was a requirement for SWiM to allow for revisions in a context as local as possible --
i.\,e.\ committing a ``fixed description'' to the CD repository instead of committing a
``new revision of a CD with `something' changed''.  SWiM acts as a browser and editor on
top of the OpenMath Subversion repository but adopts a finer granularity.  For a CD, there
is not one lengthy wiki page, but, on every request of the CD from the Subversion
repository, it is split into smaller logical units that are \emph{semantically} subject to
a revision: mathematical properties and examples on the lowest level, then symbol
definitions (grouping several mathematical properties, examples, and metadata about one
symbol together), and finally whole CDs.  Of the wiki pages on CD and symbol definition
levels, only the structural outline is editable, which keeps the content of the page
editor small and maintainable; the smaller subparts that have been split into pages of
their own right are editable separately and only represented as XInclude
links~\cite{W3C:XInclude10} in the editing view.  Nevertheless, a complete CD can be
\emph{viewed} at once; the presentation XSLTs have been adapted to cater for that.
Metadata fields are either editable within the structural outline editor, or in a separate
form-based view.  Much attention was paid to avoiding any disruption of the file
granularity of CDs in the Subversion repository, which are still editable in the
conventional way\footnote{As we will see in section~\ref{sec:experiences-directions}, SWiM
  does have, and will always have, certain technical but also conceptual limitations, be
  they bugs or deliberate design choices, that disqualify it as a one-size-fits-all CD
  editor.}.
Upon saving a change in the wiki, the whole CD to which the changed part belongs is
reassembled, reversing the initial splitting process, and committed to the repository.
However, the \emph{log message} for this commit refers to the particular part of the CD
that has been changed.  In the revision log of the CD, such a revision will display as
follows (here shown for a change of the description of the \url{transc1\#sin} symbol):

{\footnotesize
\begin{verbatim}
r1234 | clange | 2009-05-11 13:06:41 +0200 (Mon, 11 May 2009) |
 2 lines
[Administrator@SWiM] replaced metadata field dc:description
Actually changed fragment cd:transc1+sin
\end{verbatim}
}

The naming of CDs and parts thereof currently varies from OpenMath conventions and instead
reflects the SWiM-internal RDF representation (as described in the following subsection)
but could easily be adapted.  The differing user names are owed to the technical
limitation that SWiM and the Subversion repository do not have a unified user account
management.

\subsection{Discussing and Implementing Revisions}
\label{sec:swim-argu}

\begin{figure}
\centering
\includegraphics[width=\textwidth]{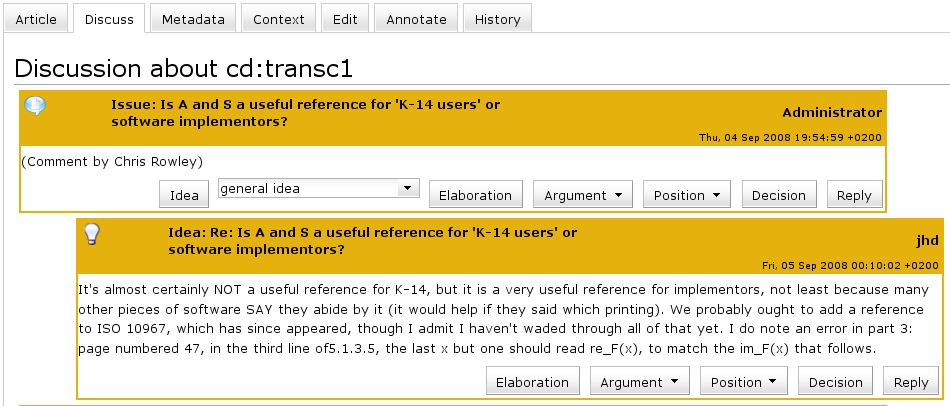}
\caption{Part of a discussion page from the OpenMath wiki.  Notice the post types and the
  specialised reply buttons.}
\label{fig:discussion}
\end{figure}

For each page (i.\,e.\ for each CD, symbol = CDDefinition, mathematical property, and
example), SWiM offers a discussion page -- essentially one local discussion forum per
subject of interest.  While that already allows discussions in the same granularity as our
units of mathematical knowledge have, we have also given the discussion threads a semantic
structure.  On a conventional wiki discussion page, users would have
to \begin{inparaenum}\item\label{it:disc-thread} manually create one section per
  discussion thread, \item\label{it:disc-reply} manually indent
  replies, \item\label{it:disc-topic} and point out the message of their discussion post
  in natural language\end{inparaenum}. The IkeWiki
platform~\cite{schaffert06:STICA-ikewiki} that SWiM is based on already cared for
(\ref{it:disc-thread}) and (\ref{it:disc-reply}) by adopting the user interface known from
discussion forums (and storing each discussion post as a separate resource instead of
storing the whole discussion page, as other wikis do).  We have added
(\ref{it:disc-topic}) in a way that optionally allows users to indicate the \emph{type} of
their discussion posts in terms of an \emph{argumentation ontology}, of which we present a
simplified outline here (see~\cite{LangeEtAl:ArgumentationSWiM08} for details): Such a
discussion can be started by pointing out a problem (here called \emph{issue}).  As
replies to an \emph{issue} post, \emph{ideas} (= solution proposals) would be allowed, on
which users can state their \emph{position}, and finally a thread can be concluded with a
post of type \emph{decision}, summarising the idea that was actually agreed on to resolve
the issue.  For every possible type of reply to a discussion post, there is a dedicated
reply button (cf.\ figure~\ref{fig:discussion}); ``untyped'' replies for posts that do not
fit into this schema are still possible but obviously prevent further automated
assistance.

Aiming at a technical support that guides discussion threads towards common solutions, we
added a domain-specific extension to the argumentation ontology.  In a survey among
OpenMath users\footnote{The survey is still open for participation at
  \url{http://tinyurl.com/5qdetd} but likely to be replaced by a more focused survey
  soon.}, patterns of common problem and solution types in mathematical knowledge bases
were identified~\cite{LangeEtAl:ArgumentationSWiM08}.  The benefit from that is twofold:
\begin{inparaenum}
\item Discussion threads can be queried by their logical structure, and
\item assistants for semi-automatically implementing common solution patterns to common
  problems can be implemented (cf.~\cite{LangeEtAl:ArgumentationSWiM08})\end{inparaenum}.  SWiM not only represents the structure of discussion threads in an RDF
graph~\cite{w3c:rdf} in terms of the above-mentioned argumentation ontology, but it also
represents the structure of CDs in terms of an ontology: part--whole links, as identified
during the splitting of CDs described in section~\ref{sec:swim-minoredit}, links from
symbol occurrences in mathematical objects to the place where they have been defined, as
well as metadata.  This whole RDF database can be queried.  On the entry page of the
OpenMath wiki, this is done in order to draw attention to unresolved issues by the
following SPARQL~\cite{PruSea08:sparql} query:

\begin{verbatim}
SELECT DISTINCT ?P WHERE {
  ?P ikewiki:hasDiscussion ?D .
  ?C a arguonto:Issue;
     sioc:has_container ?D .
  OPTIONAL { ?Dec arguonto:decides ?C . }
  FILTER (!bound(?Dec)) }
\end{verbatim}

\textit{P} is a variable for a wiki page, which could be further restricted by its type in
terms of the OpenMath ontology, e.\,g.\ we could restrict the query to symbols
(\textit{CDDefinition}).  This query returns all pages \textit{P} having a discussion
forum \textit{D} containing a comment \texttt{C} of type \textit{Issue} on which no
decision has been made so far.  Such queries can be entered anywhere by an experienced
user and result in a list of links to wiki pages.

\subsection{Editing and Verifying Notations}
\label{sec:swim-notations}

In rendering mathematical objects to Presentation MathML, SWiM adopts the approach of the
``Kohlhase camp'' (cf.\ section~\ref{sec:om3}) by embedding the JOMDoc rendering
library~\cite{TR:KLMMR:NfAD,KMR:NoLMD08} and maintaining notation dictionaries in parallel
to content dictionaries.  The notation definitions are browsable and editable in the wiki.
The workflow of editing and verifying them, as outlined in section~\ref{sec:uc-notation},
is facilitated as follows
(see~\cite{LangeGonzalez:SWiM-Sentido08,lange:swim-notation-semantics08} for details):

\begin{enumerate}
\item SWiM utilises the parallel markup~\cite[chapter 5.4]{CarlisleEd:MathML08} generated
  by the renderer to create links from the rendered symbols to the wiki pages representing
  their \textit{CDDefinition}s.  Thus, a developer can directly navigate from the
  occurrence of a symbol to its definition, and from there its notation definition is only
  one more click away.
\item The XHTML+MathML output of rendering a wiki page (= a CD or a fragment thereof) is
  cached, but after changing a notation definition of a symbol, the rendered output for
  all pages \textit{P} containing a formula in which the symbol occurs is removed from the
  cache, forcing its re-generation.  Note that the set \textit{P} contains not only the
  FMP or example that immediately holds the OpenMath object using the symbol, but also the
  enclosing CDDefinition and CD.  The set \textit{P} is obtained by another SPARQL query
  on the database.
\end{enumerate}

\section{Discussion, Experiences and Further Directions}
\label{sec:experiences-directions}

This section discusses the SWiM features presented so far, lists preliminary user feedback
about them, as well as general feedback obtained from the users of the OpenMath wiki, and
concludes with a schedule of plans for further improvement.

By supporting the use cases ``minor edits'', ``discussing and implementing revisions'' and
``editing and verifying notations'' and by its non-disruptive connection to the OpenMath
Subversion repository, SWiM facilitates crucial aspects of the CD maintenance process.
Moreover, we got a fine-grained permission system for free from the underlying IkeWiki
engine, which allows to define roles like ``visitor'' (may comment on everything), ``CD
editor'' (may edit the CDs), and ``administrator'' (may also edit special pages like the
entry page).  The OpenMath developers have made little use of the wiki for actually
\emph{changing} the CDs (for usability reasons elaborated on below), but mainly used it as
a \emph{browser} -- where is is slower but much richer in features than the statically
rendered CD presentations --, and for \emph{discussing}.

\subsection{Evaluation}
\label{sec:eval}

We have verified the principal utility of the basic argumentation ontology (without the
domain-specific extensions yet) for OpenMath by importing an old corpus of e-mail
conversations about the OpenMath/MathML 3 CDs by Chris Rowley, David Carlisle, Michael
Kohlhase, and others, into the wiki, following the discussion structure.  Further
discussion posts have been contributed by OpenMath developers afterwards.  Overall, this
resulted in 90 discussion posts.  A breakdown of this figure can be evaluated by post type
and by post granularity:
\begin{description}
\item[by type:] 69 posts fit into one of the types from the argumentation ontology, mainly
  \textit{Issue} (48) and \textit{Idea} (10).  Only counting the 23 posts contributed by
  the users themselves (who were obviously less familiar with the background of the
  argumentation ontology), the result is slightly less convincing; for 9 of them the users
  were not sure how to classify them.  The post type that was missing in most cases was
  nothing argumentative at all, but the \emph{question} -- either a direct question about
  some concept from a CD, or a follow-up question on an argumentative post, such as ``what
  do you mean by this issue description?''.  It will be easy to solve that problem by
  adding such a post type.  Some other posts could not be uniquely classified because they
  both raised an issue and proposed a solution (= idea) in the same sentence.  Annotating
  different argumentative types not at the level of posts but \emph{within} posts is
  highly non-trivial, both concerning conceptual modelling and user interface design,
  though, as discussed in~\cite{LBGBH08:SIOC-argumentation}.
\item[by granularity:] 36 posts (but only posts taken from the e-mail corpus) had
  individual symbols as their subject; the remaining 54 posts (including all of the posts
  made by users) were made on CD-level discussion pages.  This shows that either the users
  did not find it intuitive (or not necessary) to access subparts of a CD when they saw a
  complete CD in the browser, or that it was not possible to identify individual symbols a
  post referred to.  The latter is the case for certain posts that argue on design issues
  of a CD in general, sometimes naming certain individual symbols as examples.  A few
  other posts from the e-mail corpus referred to \emph{two} closely related symbols; we
  filed copies of them with both affected symbols.
\end{description}

Overall, this shows that the OpenMath CD editors have understood how to make use of this
way of discussing problems, which is more exact than writing an e-mail or opening a Trac
ticket.

The only evaluation of the editing features so far we have performed ourselves: We made
sure that no content is lost or broken from the CD files in the Subversion repository
during minor edits in SWiM.  We have tested that by importing all OpenMath 3 CDs into the
wiki, loading them into the editor once, saving them, and inspecting the XML diff.

A major criticism towards the wiki has so far been its focus on editing existing content.
The different granularities of the wiki and the OpenMath Subversion repository make it
very cumbersome to add, e.\,g., a new symbol to a CD: One has to edit the CD wiki page,
add the new \textit{CDDefinition} child there, as a sibling of the XInclude elements
pointing to the existing \textit{CDDefinition}s, and then save the CD page.  Upon saving,
the new \textit{CDDefinition} fragment will be split away into a wiki page of its own,
which can then be edited in the next step.  Cleanly adding a new CD altogether is not
possible at all, this time due to the incomplete Subversion support of SWiM.  SWiM only
implements the most basic Subversion commands so far: \texttt{update}, \texttt{commit},
and \texttt{lock}.  Other actions like adding and deleting content are possible in the
wiki itself but not reflected by its interface to Subversion -- which is hacked into the
file import/export component instead of being integrated at database level, because the
latter would have required a complete overhaul of the design of the underlying IkeWiki
system.

\subsection{Roadmap}
\label{sec:roadmap}

These and other annoyances and missing features (not being able to link to discussion
posts, no e-mail notification about discussion posts or page changes, no global
search/replace feature across multiple symbols or CDs, to name just a few) are hard to
resolve within the existing architecture of SWiM.  While some major tasks are definitely
within the responsibility of the author, the general usability of the system -- besides
its adaptation to the mathematical domain -- could benefit a lot from improvements to the
underlying wiki engine.  The development of IkeWiki, which had originally been chosen due
to its unique XML and RDF support, has been discontinued, though.  On the other hand, its
completely reengineered successor KiWi~\cite{SchaffertElAl:KiWi09} is making good
progress.

Therefore, a port of SWiM to KiWi is currently in progress.  KiWi's more modular
architecture allows for implementing large parts of SWiM not by modifying the core system
-- as was the case with IkeWiki --, but by providing plugins.  New KiWi features of
particular interest in the OpenMath setting are a dashboard view giving every user a
personalised overview of recent changes at a glance, a service that recommends related
content, a facetted search interface, and a concept of transactions that will allow for
committing several related changes at once.  With the new, improved SWiM system, we will
then restart the usability evaluation and work out an accompanying user questionnaire.

A further enhancement planned is replacing the wiki's own database by an integration of
Subversion on database level.  A database engine capable of versioning XML documents,
particularly mathematical documents, is currently under development in our
group~\cite{TNTBase:web}.  On the user interface end, it is planned to make the OpenMath
community benefit from our recent research on active documents.  We have implemented
interactive services like in-place definition lookup and developed an infrastructure for
user-adaptable documents~\cite{GLR:WebSvcActMathDoc09}.

\subsection{Conclusion}
\label{sec:conc}

We have outlined three CD editing use cases and compared the traditional way of performing
them to the new way offered by the SWiM wiki.  SWiM clearly excels in these special but
common use cases, which has partly been confirmed by the OpenMath CD editors, while still
staying compatible with old-style operations going on in the same repository.  As SWiM
does not yet cover the full CD editing workflow, we presented a roadmap towards its
successor, which will rely on a smarter database backend and increase the interactivity of
the \url{http://wiki.openmath.org} site for current and future collaborators and users.

\paragraph{Acknowledgments:} The author would like to thank the members of the OpenMath
Society, particularly (in alphabetical order) Olga Caprotti, David Carlisle, James
Davenport, Paul Libbrecht, Michael Kohlhase, Jan Willem Knopper, and Chris Rowley for
giving helpful hints and testing during the design and setup of the wiki, Alberto González
Palomo for developing the Sentido formula editor employed by SWiM, and Jakob Ücker for
carrying out most of the evaluation work.  This work was supported by JEM-Thematic-Network
ECP-038208.

\bibliographystyle{abbrv}
\bibliography{kwarc}

\providecommand\seen{seen } \providecommand\webpageat{web page at }
  \providecommand\homepageat{home page at }
  \providecommand\projectpageat{project page at }
  \providecommand\systempageat{system home page at }
  \providecommand\svnrepoat{Subversion repository at }
  \providecommand\January{January} \providecommand\February{February}
  \providecommand\March{March} \providecommand\MARCH{March}
  \providecommand\April{April} \providecommand\May{May}
  \providecommand\June{June} \providecommand\JUNE{June}
  \providecommand\July{July} \providecommand\JULY{July}
  \providecommand\August{August} \providecommand\September{September}
  \providecommand\October{October} \providecommand\November{November}
  \providecommand\December{December} \providecommand\AUSTRALIA{Australia}
  \providecommand\ROMANIA{Romania} \providecommand\MEXICO{Mexico}
  \providecommand\ITALY{Italy} \providecommand\USA{USA}
  \providecommand\IRELAND{Ireland} \providecommand\HUNGARY{Hungary}
  \providecommand\JAPAN{Japan} \providecommand\CANADA{Canada}
  \providecommand\SPAIN{Spain} \providecommand\NETHERLANDS{Netherlands}
  \providecommand\UK{UK} \providecommand\SWEDEN{Sweden}
  \providecommand\GERMANY{Germany} \providecommand\germany{Germany}
  \providecommand\Germany{Germany} \providecommand\openmath{OpenMath}
  \providecommand\fc{forthcoming} \providecommand\PROC{Proceedings}
  \providecommand\omdoc{OMDoc} \providecommand\activemath{ActiveMath}
\begin{thebibliography}{10}

\bibitem{CarlisleEd:MathML08}
R.~Ausbrooks, B.~Bos, O.~Caprotti, D.~Carlisle, G.~Chavchanidze, A.~Coorg,
  S.~Dalmas, S.~Devitt, S.~Dooley, M.~Hinchcliffe, P.~Ion, M.~Kohlhase,
  A.~Lazrek, D.~Leas, P.~Libbrecht, M.~Mavrikis, B.~Miller, R.~Miner,
  M.~Sargent, K.~Siegrist, N.~Soiffer, S.~Watt, and M.~Zergaoui.
\newblock {M}athematical {M}arkup {L}anguage ({MathML}) version 3.0.
\newblock {W3C} working draft of november 17., World Wide Web Consortium, 2008.

\bibitem{MKM09}
J.~Carette, L.~Dixon, C.~{Sacerdoti Coen}, and S.~M. Watt, editors.
\newblock {\em {MKM/Calculemus} 2009 Proceedings}, number 5625 in LNAI.
  Springer Verlag, 2009.
\newblock in Press.

\bibitem{Davenport:stso99}
J.~Davenport.
\newblock A small {\openmath} type system.
\newblock Technical report, The {\openmath} Esprit Project, 1999.

\bibitem{DavKoh:umoattsea09}
J.~H. Davenport and M.~Kohlhase.
\newblock {Unifying Math Ontologies: A tale of two standards}.
\newblock In Carette et~al. \cite{MKM09}.
\newblock in Press.

\bibitem{JDPL:TFtEOaiU}
J.~H. Davenport and P.~Libbrecht.
\newblock {The Freedom to Extend {\openmath} and its Utility}.
\newblock {\em Journal of Mathematics and Computer Science, special issue on
  Mathematical Knowledge Management}, 2008.

\bibitem{GLR:WebSvcActMathDoc09}
J.~Giceva, C.~Lange, and F.~Rabe.
\newblock Integrating web services into active mathematical documents.
\newblock In Carette et~al. \cite{MKM09}, pages 279--293.
\newblock in Press.

\bibitem{TR:KLMMR:NfAD}
M.~Kohlhase, C.~Lange, C.~M{\"u}ller, N.~M{\"u}ller, and F.~Rabe.
\newblock Notations for active mathematical documents.
\newblock {KWARC} Report 2009-1, Jacobs University Bremen, 2009.

\bibitem{KohLanRab:pmcfe07}
M.~Kohlhase, C.~Lange, and F.~Rabe.
\newblock Presenting mathematical content with flexible elisions.
\newblock In O.~Caprotti, M.~Kohlhase, and P.~Libbrecht, editors, {\em
  {\openmath}/ JEM Workshop 2007}, 2007.

\bibitem{KMR:NoLMD08}
M.~Kohlhase, C.~M{\"u}ller, and F.~Rabe.
\newblock Notations for living mathematical documents.
\newblock In S.~Autexier, J.~Campbell, J.~Rubio, V.~Sorge, M.~Suzuki, and
  F.~Wiedijk, editors, {\em {Intelligent Computer Mathematics}, 9th
  International Conference, AISC 2008 15th Symposium, Calculemus 2008 7th
  International Conference, MKM 2008 Birmingham, UK, July 28 - August 1, 2008,
  Proceedings}, number 5144 in LNAI, pages 504--519. Springer Verlag, 2008.

\bibitem{Lange:SWiM-OpenMath08}
C.~Lange.
\newblock Editing {\openmath} content dictionaries with swim.
\newblock In {\em 3rd JEM Workshop (Joining Educational Mathematics)}, 2008.

\bibitem{lange:swim-notation-semantics08}
C.~Lange.
\newblock Mathematical semantic markup in a wiki: The roles of symbols and
  notations.
\newblock In C.~Lange, S.~Schaffert, H.~Skaf-Molli, and M.~V{\"o}lkel, editors,
  {\em Proceedings of the 3\textsuperscript{rd} Workshop on Semantic Wikis,
  {European} {Semantic} {Web} {Conference} 2008}, volume 360 of {\em CEUR
  Workshop Proceedings}, Costa Adeje, Tenerife, Spain, June 2008.

\bibitem{lange:swim-demo08}
C.~Lange.
\newblock {SWiM} -- a semantic wiki for mathematical knowledge management.
\newblock In S.~Bechhofer, M.~Hauswirth, J.~Hoffmann, and M.~Koubarakis,
  editors, {\em ESWC}, volume 5021 of {\em Lecture Notes in Computer Science},
  pages 832--837. Springer, 2008.

\bibitem{LBGBH08:SIOC-argumentation}
C.~Lange, U.~Boj{\=a}rs, T.~Groza, J.~Breslin, and S.~Handschuh.
\newblock Expressing argumentative discussions in social media sites.
\newblock In J.~Breslin, U.~Boj{\=a}rs, A.~Passant, and S.~Fern{\'a}ndez,
  editors, {\em Social Data on the Web (SDoW2008), Workshop at the 7th
  International Semantic Web Conference}, Oct. 2008.

\bibitem{LangeGonzalez:SWiM-Sentido08}
C.~Lange and A.~Gonz{\'a}lez~Palomo.
\newblock Easily editing and browsing complex {OpenMath} markup with {SWiM}.
\newblock In P.~Libbrecht, editor, {\em Mathematical User Interfaces Workshop
  2008}, 2008.

\bibitem{LangeEtAl:ArgumentationSWiM08}
C.~Lange, T.~Hastrup, and S.~Corlosquet.
\newblock Arguing on issues with mathematical knowledge items in a semantic
  wiki.
\newblock In J.~Baumeister and M.~Atzm{\"u}ller, editors, {\em {Wissens- und
  Erfahrungsmanagement LWA (Lernen, Wissensentdeckung und Adaptivit{\"a}t)
  Conference Proceedings}}, volume 448, 2008.

\bibitem{W3C:XInclude10}
J.~Marsh, D.~Orchard, and D.~Veillard.
\newblock {XML} inclusions ({XInclude}) version 1.0 (second edition).
\newblock {W3C Recommendation}, World Wide Web Consortium ({W3C}), Nov. 2006.

\bibitem{Openmath:web}
{{\openmath} Home}.
\newblock {\url{http://www.openmath.org/}}, seen May 2009.

\bibitem{PruSea08:sparql}
E.~Prud'hommeaux and A.~Seaborne.
\newblock {SPARQL} query language for {RDF}.
\newblock {W3C} {R}ecommendation, World Wide Web Consortium, Jan. 2008.

\bibitem{w3c:rdf}
Resource description framework ({RDF}).
\newblock \url{http://www.w3.org/RDF/}, 2004.

\bibitem{URL:riaca-openmath}
{RIACA OpenMath products}.
\newblock \webpageat\url{http://www.riaca.win.tue.nl/projects/openmath/}.

\bibitem{schaffert06:STICA-ikewiki}
S.~Schaffert.
\newblock {IkeWiki}: A semantic wiki for collaborative knowledge management.
\newblock In {\em 1\textsuperscript{st} International Workshop on Semantic
  Technologies in Collaborative Applications STICA 06, Manchester, UK}, June
  2006.

\bibitem{SchaffertElAl:KiWi09}
S.~Schaffert, J.~Eder, S.~Gr{\"u}nwald, T.~Kurz, M.~Radulescu, R.~Sint, and
  S.~Stroka.
\newblock {KiWi} -- a platform for semantic social software.
\newblock In C.~Lange, S.~Schaffert, H.~Skaf-Molli, and M.~V{\"o}lkel, editors,
  {\em Proceedings of the 4\textsuperscript{th} Workshop on Semantic Wikis,
  {European} {Semantic} {Web} {Conference} 2009}, Hersonissos, Greece, June
  2009.
\newblock in press.

\bibitem{TNTBase:web}
{TNTBase} {Project}, seen December 2008.
\newblock available at \url{https://trac.mathweb.org/tntbase/}.

\bibitem{trac:web}
{The Trac Project}.
\newblock \url{http://trac.edgewall.org/}, 2008.

\end{thebibliography}


\ifdraft
\ednotemessage
\fi
\end{document}